\documentclass[twocolumn,
preprintnumbers,amsmath,amssymb,prl]{revtex4-1}

\usepackage{amsmath,amssymb,,epsfig}

\usepackage{CJK}

\usepackage{dcolumn}

\def\be{\begin{eqnarray}}
\def\ee{\end{eqnarray}}
\def\bea{\begin{eqnarray}}
\def\eea{\end{eqnarray}}

\def\half{\frac{1}{2}}

\begin{document}

\preprint{YITP-SB-11-13}

\begin{CJK*}{GB}{}
\CJKfamily{gbsn}

\title{
 The $4d$ Superconformal Index from $q$-deformed $2d$ Yang-Mills
}
\author{Abhijit Gadde}
\author{Leonardo Rastelli}
\author{Shlomo S. Razamat}
\author{Wenbin Yan (ÑÕÎıó)}

\affiliation{ C.~N.~Yang institute for theoretical physics\\
Stony Brook University\\
Stony Brook, NY 11794 USA\\}

\date{\today}

\begin{abstract}

We  identify   the $2d$ topological theory underlying
the ${\cal N}=2$ $4d$ superconformal index with an explicit model:  $q$-deformed $2d$  Yang-Mills.
By this route we  are able to  evaluate the index of  some strongly-coupled $4d$ SCFTs,
such as Gaiotto's $T_N$ theories.

\end{abstract}


\maketitle

\end{CJK*}

\section{Introduction}

In this letter we  describe a new powerful duality, relating physics in four and in two dimensions.
 We will argue that for a large class of \textit{four-dimensional  superconformal} gauge theories,
 non-trivial information about the operator spectrum is  captured by correlators of a \textit{two-dimensional non-supersymmetric} gauge theory.
 The $4d$ side of the duality is generically   strongly-coupled, and difficult to analyze directly;
 on the other hand calculations on the $2d$ side will be  explicit and algorithmic.
  Thus our conjecture  gives new information about strongly-coupled $4d$ field theories.

 Our proposal is in the same spirit as the Alday-Gaiotto-Tachikawa (AGT)
 relation between the  partition function of  a $4d$ ${\cal N}=2$ gauge theory on $S^4$
and a correlator in $2d$ Liouville/Toda
theory~\cite{Alday:2009aq,Wyllard}. In our case, the $4d$ observable
is a (twisted)
 supersymmetric partition function  of an ${\cal N}=2$ superconformal field theory on $S^3 \times S^1$,
also known as the superconformal index. We will focus on a ``reduced'' index that depends
 on a single fugacity $q$.
On the $2d$ side, instead of   Liouville/Toda we have the {\it
zero-area limit} of   $q$-deformed Yang-Mills
 theory.
The topological nature of this $2d$ theory dovetails with the independence of the $4d$ index on the gauge theory moduli.

We begin by reviewing the $4d$ side of the duality.
The full ${\cal N} =2$ superconformal index is defined as~\cite{Kinney:2005ej}
 \be
 \mathcal{I}=  \mbox{Tr}(-1)^{F}p^{\frac{E-R}{2}+j_1}q^{\frac{E-R}{2}-j_1}u^{-(r+R)} \,,
 \ee
where the trace is  over the states of the theory on $S^3$ (in the usual radial quantization) and $F$ the fermion number.
The symbol $E$ stands for the conformal dimension, $(j_1, j_2)$ for the Cartan generators of the $SU(2)_1 \otimes SU(2)_2$   isometry group, and $(R \, ,r)$ for the Cartan generators
 of the  $SU(2)_R \otimes U(1)_r$ R-symmetry.
The fugacities $p$, $q$, and $u$
 keep track of the maximal set of quantum numbers
 commuting with a single real supercharge, ${\cal Q} \equiv \tilde {\cal Q}_{1 \dot -}$, which with no loss of generality has been chosen to have $R=\frac{1}{2} $, $r=-\frac{1}{2}$, $j_1=0$, $j_2 =-\frac{1}{2}$
and (of course)  $E=\frac{1}{2}$.
Only states that obey $ 2\{ {\cal Q}, {\cal Q}^\dagger \} =  E - 2 j_2 - 2 R +r = 0$ contribute to the index. Note that
the variables   $p$, $q$, and $u$ are related to $t,y,v$ of \cite{Gadde:2009kb} as $p=t^3y,\,q=\frac{t^3}{y}$ and $u=\frac{v}{t}$.

 For a theory with a weakly-coupled Lagrangian description the index is computed explicitly by a matrix integral,
\be
\label{index}
&&{\cal I}(p,q,u;V) =\int\left[dU\right]\,\\
&&\exp\left(\sum_{n=1}^\infty\frac{1}{n}\;\sum_{j} f^{(j)}(p^n,q^n,u^n)  \, \chi_{{\mathcal R_j}}(U^n,\,V^n)\right) \, .\nonumber
\ee Here $U$ denotes an element of the gauge group,  with $\left[dU\right]$  the invariant Haar measure,
and $V$ an element   of the flavor group. The  sum is over the different ${\cal N}=2$ supermultiplets appearing in the Lagrangian,
with  ${\mathcal R_j}$ the representation of the $j$-th multiplet under the flavor and  gauge groups  and $\chi_{\mathcal R_j}$  the corresponding character.
The functions $f^{(j)}$ are the ``single-letter'' partition functions, $f^{(j)}= f^{vect}$ or $f^{(j)} = f^{chi}$ according to whether the $j$-th multiplet
is an ${\cal N}=2$ vector or ${\cal N}=2$ $\frac{1}{2}$-hypermultiplet.
  They are easily evaluated \cite{Kinney:2005ej}:
\be\label{letterpart}
&&  f^{vect}(p,q,u)=\frac{(u-\frac{1}{u})\sqrt{pq}-(p+q)+2pq}{(1-p)(1-q)}\,,\\
&&f^{chi}(p,q,u)=\frac{(pq)^{\frac{1}{4}}\frac{1}{\sqrt{u}}-(pq)^{\frac{3}{4}}\sqrt{u}}{(1-p)(1-q)}\,.
\ee
We will focus on a \textit{reduced} index,
by setting
\be
u = 1,\qquad p=q \, ,
\ee
which leads to the significant simplification
\be
f^{vect}=\frac{-2q}{1-q},\qquad\qquad  f^{chi}=\frac{q^{\frac{1}{2}}}{1-q}\,.
\ee
We consider a
class of  ${\cal N}=2$ $4d$ superconformal theories (SCFTs) constructed from a set of elementary building blocks~\cite{Gaiotto:2009we}. The building blocks are
isolated SCFTs with flavor symmetry $G_1 \otimes G_2 \otimes G_3$,
$G_i\subseteq SU(N)$ for given  $N$.
In the simplest case of $N=2$,  the only building
block is the free $\frac{1}{2}$-hypermultiplet in the tri-fundamental representation
of the $SU(2)^3$ flavor group. For $N  >2$ most of the building blocks are intrinsically strongly-interacting theories with no
 Lagrangian description. One can ``glue together'' two building blocks by gauging a common $SU(N)$ flavor symmetry.
Iterating this procedure one constructs
a large class of  ${\mathcal N}=2$ gauge theories, the $SU(N)$ ``generalized quivers'' ~\cite{Gaiotto:2009we}.
There is a geometric
interpretation of  this construction, where one regards the building blocks as three-punctured spheres, with the punctures associated
to the flavor symmetries; the gluing operation is performed by connecting the punctures
with cylinders. The complex structure moduli of the resulting punctured Riemann surface correspond to the complexified gauge couplings.
The same punctured Riemann surface can often
be obtained by following several different gluing paths (different
pairs-of-pants decompositions). The generalized quiver theories associated
to different decompositions of the same surface are  related by S-dualities ~\cite{Gaiotto:2009we}.

The index of a generalized quiver can be written in terms
 of the index of its constituents.
We parametrize the index of an elementary building block (3-punctured sphere) by ``structure constants''
$\mathcal{I}_N(\mathbf{x_1},\mathbf{x_2},\mathbf{x_3})$ where $\mathbf{x}_i$ are fugacities dual to the Cartan subgroup
of $G_i$: except in special cases these are {\it a priori} unknown functions.
 On the other hand we can easily write the index  $\eta_N({\mathbf x})$ of the $SU(N)$ vector multiplets used in the gluing (propagators),
\be
 \eta_N({\mathbf x})=\exp\left[
-2\sum_{n=1}^\infty\frac{1}{n}\frac{q^{n}}{1-q^{n}}\chi_{adj}(\mathbf{x}^n)\right]\, .\nonumber
\ee
 For example, gluing two 3-punctured spheres with one cylinder one obtains the
following index
\be \label{following}
\int[dU({\mathbf x})]\,\mathcal{I}_N({\mathbf x_1},{\mathbf x_2},{\mathbf x})\,\eta_N({\mathbf x})\,\mathcal{I}_N({\mathbf x},{\mathbf x_3},{\mathbf x_4})\,.
\ee By defining a metric
\be
\eta_N({\mathbf x_1},{\mathbf x_2}) \equiv \eta_N({\mathbf x_1})\,\sum_{{\mathcal R}}
 \chi_{{\mathcal R}}({\mathbf x_1})\,\chi_{{\mathcal R}}({\mathbf x_2})\,,
\ee with ${\mathcal R}$ running over irreducible and finite representations of
$SU(N)$, we can re-write (\ref{following}) as
\be
 \mathcal{I}_N(\mathbf{x_1},\mathbf{x_2},\mathbf{x})\cdot\,\eta_N({\mathbf x},{\mathbf x'})\cdot\,\mathcal{I}_N(\mathbf{x'},\mathbf{x_3},\mathbf{x_4})\,,
\ee where $\cdot$ multiplication means  integration over the Haar measure. S-duality then implies that the metric and structure
constants
form an associative algebra and thus  a $2d$
 topological field theory (TQFT)~\cite{Gadde:2009kb}. (Strictly speaking, the state-space at each puncture,
which is spanned by $G_i$ representations, is infinite-dimensional,
 so one must slightly relax the standard mathematical axioms for a TQFT.)
Associativity was directly verified for the $SU(2)$ and $SU(3)$ generalized quiver theories in~\cite{Gadde:2009kb,Gadde:2010te},
for generic values of the fugacities $p,\,q$ and $u$.
 In the following we will identify the $2d$ topological theory implicitly defined by the reduced index with
an explicit model:   $q$-deformed  Yang-Mills ($q$YM) in the zero-area limit.

\section{$SU(2)$ generalized quivers}
Let us start with the simplest case, the $SU(2)$ quivers.
Here the building blocks are free tri-fundamental $\frac{1}{2}$-hypermultiplets,
\be
&&\mathcal{I}_{222}(a_1,a_2,a_3)=\exp\biggl[\sum_{n=1}^\infty\frac{1}{n}\frac{q^{\frac{1}{2}n}}{1-q^{n}}\chi_\square(a_1^n)\chi_\square(a_2^n)\chi_\square(a_3^n)\biggr]\,.\nonumber
\ee
Remarkably, one can prove ({\it e.g.} by comparing analytic properties) that ${\mathcal I}_{222}(a_1,a_2,a_3)$ admits  the equivalent representation
\be\label{su2norm}
&& \mathcal{I}_{222}(a_1,a_2,a_3) =\\
&&\qquad  \frac{(q;q)_\infty}{1-q} \, \prod_{i=1}^{3}  \eta_2^{-\half}(a_i)
 \sum_{{\mathcal R}} \frac{\chi_{{\mathcal R}}(a_1)\,
\chi_{{\mathcal R}}(a_2)\,\chi_{{\mathcal R}}(a_3)}{\left[|{\mathcal R}|\right]_q}\,.\nonumber
\ee
Here $(q;q)_\infty\equiv \prod_{i=1}^\infty(1-q^{i})$.
The sum  is over irreducible $SU(2)$ representations ${\mathcal R}$,
with $|{\mathcal R}|$ denoting  the dimension of the representation.
The $SU(2)$ characters are
\be
\chi_{{\mathcal R}}(a)=\frac{a^{|{\mathcal R}|}-a^{-|{\mathcal R}|}}{a-a^{-1}}\,.
\ee
Finally the symbol  $[x]_q$ denotes the \textit{q-deformed} number,
\be
[x]_q \equiv \frac{q^{-\frac{x}{2}}-q^{\frac{x}{2}}}{q^{-\frac{1}{2}}-q^{\frac{1}{2}}}\,.
\ee
The structure constants contain the factors $\prod_i \eta_2^{-1/2}(a_i)$, which cancel with the metric $\eta_2(a_i)$
when two punctures are glued. It is then
natural to define
 rescaled structure constants and metric,
\be
 \hat {\mathcal I}_{222}(a_1,a_2,a_3)&=& {\cal N}_{222}(q) \sum_{{\mathcal R}} \frac{\chi_{{\mathcal R}}(a_1)\,
\chi_{{\mathcal R}}(a_2)\,\chi_{{\mathcal R}}(a_3)}{\left[|{\mathcal R}|\right]_q}\,,\nonumber
\\
\hat\eta_2(a,\,b)&=&\sum_{{\mathcal R}} \chi_{{\mathcal R}}(a)\,\chi_{{\mathcal R}}(b)\, ,
\ee
where  ${\cal N}_{222}(q) =  (q;q)_\infty /(1-q)$.
Up to the overall normalization ${\cal N}_{222}$, these are precisely the structure constants and metric of  $2d$ $q$YM in the zero area limit~\cite{Aganagic:2004js, Buffenoir:1994fh}!

Note that $[n]_q=\chi_n(q^{1/2})$. This  implies that by setting one of the $SU(2)$ fugacities to $q^{1/2}$ we ``close'' a
puncture,
\be
\hat {\mathcal I}_{222}(a,b,q^{1/2})&=&
{\cal N}_{222}(q) \;\; \hat\eta_2(a,\,b)\,.\nonumber
\ee
Applying this procedure again, we close another puncture and obtain the one-punctured sphere (the cap).
For higher-rank groups we will encounter  a similar procedure: setting
some combination of the flavor  fugacities  to $q^{1/2}$ one obtains punctures with reduced flavor symmetry.

\section{$SU(3)$ generalized quivers}

Next let us consider the $SU(3)$ generalized quivers. Here two new generic features appear. First, the basic building block
is an interacting theory with no Lagrangian description, the $E_6$ SCFT~\cite{Argyres:2007cn,Gaiotto:2009we}.
Second, there is more than one type of puncture: in addition
to the \textit{maximal} $SU(3)$ flavor puncture there is a puncture with reduced flavor symmetry, $U(1)$ ~\cite{Gaiotto:2009we}.

The representations of $SU(N)$ are parametrized by $N$ integers
$\lambda_1\geq\lambda_2...\geq\lambda_{N-1}\geq \lambda_N=0$, the
row lengths of the corresponding Young diagram. The $q$-deformed
dimension of the representation is \be {\rm dim}_q{\mathcal
R}_{\underline{\lambda}}=\prod_{i<j}\frac{[\lambda_i-\lambda_j+j-i]_q}{[j-i]_q}\,,
\ee and the characters are given by Schur polynomials \be
\chi_{\underline{\lambda}}(\mathbf{x})=\frac{\det\left({x_i}^{\lambda_j+k-j}\right)}{\det\left({x_i}^{k-j}\right)}\,.
\ee Specializing to $SU(3)$ we can parametrize all the Young
diagrams by $(\lambda_1,\lambda_2)$. We observe again that the
$q$-dimension of a representation is equal to the group character
with a particular choice of fugacities, \be
\chi_{\lambda_1,\lambda_2}(q,1,q^{-1})={\rm dim}_q {\mathcal
R}_{\lambda_1,\lambda_2}\,. \ee

\subsection{Three Maximal Punctures }

The sphere with three maximal punctures corresponds to the strongly coupled
$E_6$ SCFT (the $SU(3)^3$ flavor symmetry is accidentally enhanced to $E_6$.) This theory has no Lagrangian description and thus we do not have a direct
way to compute its index. However, this index was computed~\cite{Gadde:2010te} indirectly by employing
Argyres-Seiberg
duality~\cite{Argyres:2007cn}.
Inspired by the $SU(2)$ case, we conjecture that  the index $\mathcal{I}_{E_6}(\{\mathbf{x}_i\}_{i=1}^3)$ of the $E_6$ SCFT
is proportional to the structure constants  $C_{SU(3)_q}$ of $q$-deformed $SU(3)$ Yang-Mills,
\be
&&\mathcal{I}_{E_6}( \mathbf{x_i} )=
\left[\prod_{i=1}^3\eta^{-\half}({\mathbf x_i})\right]   {\cal N}_{333} (q)\, C_{SU(3)_q} ( \mathbf{x_i} )\,,\nonumber
\ee where
\be
 C_{SU(3)_q}( \mathbf{x_i} )= \sum_{0\leq\lambda_2\leq\lambda_1}^\infty
 \frac{\chi_{\lambda_1,\lambda_2}(\mathbf{x}_1)\chi_{\lambda_1,\lambda_2}(\mathbf{x}_2)\chi_{\lambda_1,\lambda_2}(\mathbf{x}_3)}{{\rm dim}_q{\mathcal R}_{\lambda_1,\lambda_2}},\nonumber
\ee
and ${\cal N}_{333} (q)$ a normalization factor. Using {\it Mathematica},
we have checked this proposal against the results of~\cite{Gadde:2010te} to several orders in $q$, and in the process
determined the normalization to be
\be
{\cal N}_{333}(q)= \frac{{(q;q)^2_\infty}}{(1-q)^2(1-q^2)} \,.
\ee
\subsection{Two Maximal and One $U(1)$ Puncture}
Another building block is given by a sphere with two $SU(3)$ punctures and one $U(1)$ puncture.
This corresponds to a free hypermultiplet in the bi-fundamental
of $SU(3)^2$ and charged under the $U(1)$.
The index of this theory is explicitly given by
\be
&&{\mathcal I}_{331}(\mathbf{x_1},\mathbf{x_2};\;a)=
\exp\biggl[\sum_{n=1}^\infty\frac{1}{n}\frac{q^{\frac{1}{2}n}}{1-q^{n}}\chi_{hyp}(\mathbf{x_1}^n,\mathbf{x_2}^n;\;a^n)\biggr]\,,\nonumber
\ee where the flavor character is
\be
\chi_{hyp}(\mathbf{x_1},\mathbf{x_2};\;a)=\sum_{i,j} (x^i_1x_2^j a+\frac{1}{x^i_1x_2^j a})\,.
\ee
One can verify by series expansion in $q$ that
\be
&&{\mathcal I}_{331}(\mathbf{x_1},\mathbf{x_2};\;a)=C_{SU(3)_q}(\mathbf{x_1},\mathbf{x_2};\;a)\times\\
&&\qquad\frac{\prod_{i=1}^2\eta^{-\half}({\mathbf x_i})}{\prod_{\ell=1}^2(1-q^\ell)}\;
\exp\biggl[\sum_{n=1}^\infty\frac{q^{\frac{3}{2}n}}{1-q^{n}}\frac{a^{3n}+a^{-3n}}{n}\biggr]\,,\nonumber
\ee with
\be
&&C_{SU(3)_q}(\mathbf{x_1},\mathbf{x_2};\;a)=\\
&&\sum_{0\leq\lambda_2\leq\lambda_1}^\infty
\frac{\chi_{\lambda_1,\lambda_2}(\mathbf{x}_1)\,
\chi_{\lambda_1,\lambda_2}(\mathbf{x}_2)\,\chi_{\lambda_1,\lambda_2}(a\,q^{1/2},aq^{-1/2},a^{-2})}{{\rm
dim}_q{\mathcal R}_{\lambda_1,\lambda_2}}\,.\nonumber \ee Note that
this result can be recovered by starting from the structure constant
with maximal punctures and ``partially closing'' one of the
punctures by embedding $SU(2)$ fugacities
$(q^{\frac{1}{2}},q^{-\frac{1}{2}})$ into fugacities of $SU(3)$.

\section{General statement}
The generic building block of a higher-rank quiver is an interacting SCFT with no Lagrangian description.
Unlike the case of $SU(2)$ and $SU(3)$ quivers it is very hard to calculate the index of these theories, either directly or indirectly.
 However, we can naturally extrapolate the relation to $2d$ $q$YM
 to higher-rank groups.

We conjecture that the reduced index of the theory corresponding to sphere with three maximal punctures (the $T_{N}$ theory  of~\cite{Gaiotto:2009we}) is
\be \label{conjecture}
{\mathcal I}_{T_{N}}(\mathbf{x_i})
=\frac{{(q;q)^{N-1}_\infty}\prod_{i=1}^3\eta^{-\half}({\mathbf x_i})}{\prod_{\ell=1}^{N-1}(1-q^\ell)^{N-\ell}}
\; C_{SU(N)_q}(\mathbf{x_i})\,
\nonumber
\ee where
\be
&&C_{SU(N)_q}(\mathbf{x_i})=\sum_{{\mathcal R}}
  \,\frac{1}{{\rm dim}_q{\mathcal R}}\,
\chi_{{\mathcal R}}(\mathbf{x}_1)\,\chi_{{\mathcal R}}(\mathbf{x}_2)\,\chi_{{\mathcal R}}(\mathbf{x}_3)\nonumber
\ee are the structure constant of $SU(N)$ $q$YM. The sum is over irreducible $SU(N)$ representations and $\{ \mathbf{x}_i \}$
are the fugacities dual to  the Cartan subgroup.

This conjecture can be tested against the numerous S-dualities of the generalized quivers~\cite{Gaiotto:2009we}.
For instance, a linear superconformal quiver theory with two $SU(4)$ nodes admits a dual description in terms of  $T_4$
 coupled to $SU(3)$ gauge theory which in turn is coupled to an $SU(2)$ gauge theory with a single hypermultiplet.
 We have checked, in the $q$ expansion, that the indices on both sides of the duality indeed match if one uses our conjecture for the $T_4$ index.

 Another test is to compare with  physical expectations for the spectrum of protected
operators. A class of protected operators in the $T_N$ theories are the Higgs branch operators~\cite{Benini:2009mz}.
These come in  two families: $E=2,\, R=1$ in flavor representation $(adj,1,1)\oplus(1,adj,1)\oplus(1,1,adj)$ and $E=N-1, R=\frac{N-1}{2}$
in representation $(N,N,N)\oplus(\bar N,\bar N,\bar N)$.
It is straightforward to see that these operators appear in our conjecture  for the index:
the first family comes from the $\eta(\mathbf{x})^{-\half}$ factors, and the second from the
$\chi_{\square}(\mathbf{x}_1)\chi_{\square}(\mathbf{x}_2)\chi_{\square}(\mathbf{x}_3)$ and
$\chi_{\overline\square}(\mathbf{x}_1)\chi_{\overline\square}(\mathbf{x}_2)\chi_{\overline\square}(\mathbf{x}_3)$ terms in $C_{SU(N)_q}$.

We can generalize the conjecture to the structure constants with two maximal punctures and one
$U(1)$ puncture,
\be
&& {\mathcal I}_{N N1}(\mathbf{x_1}, \mathbf{x_2}, a)= \exp\biggl[\sum_{n=1}^\infty\frac{1}{n}\frac{q^{\frac{1}{2}n}}{1-q^{n}}\chi_{hyp}(\mathbf{x_1}^n,\mathbf{x_2}^n;\;a^n)\biggr]=
\nonumber\\
&&\frac{C_{ {SU(N)}_q}(\mathbf{x_1},\mathbf{x_2};\;a)}{\prod_{i=1}^2\eta^{\half}({\mathbf x_i})\prod_{\ell=1}^{N-1}(1-q^\ell)}
\exp\biggl[\sum_{n=1}^\infty\frac{q^{\frac{N}{2}n}}{1-q^{n}}\frac{a^{N n}+a^{-N n}}{n}\biggr]\, ,\nonumber
\ee
where
structure constants $C_{{SU(N)}_q}(\mathbf{x_1},\mathbf{x_2};a)$ are
\be
&&C_{SU(N)_q}(\mathbf{x_1},\mathbf{x_2};a)=\\
&&\sum_{{\mathcal R}}
  \,\frac{1}{{\rm dim}_q{\mathcal R}}\,
\chi_{{\mathcal R}}(\mathbf{x}_1)\,\chi_{{\mathcal R}}(\mathbf{x}_2)\,\chi_{{\mathcal R}}(a q^{\frac{N-2}{2}},..,a q^{-\frac{N-2}{2}},a^{1-N})\,.\nonumber
\ee
Again we have verified this conjecture in the $q$-expansion.
\begin{figure}[htbp]
\begin{center}
\centerline{\includegraphics[scale=0.35]{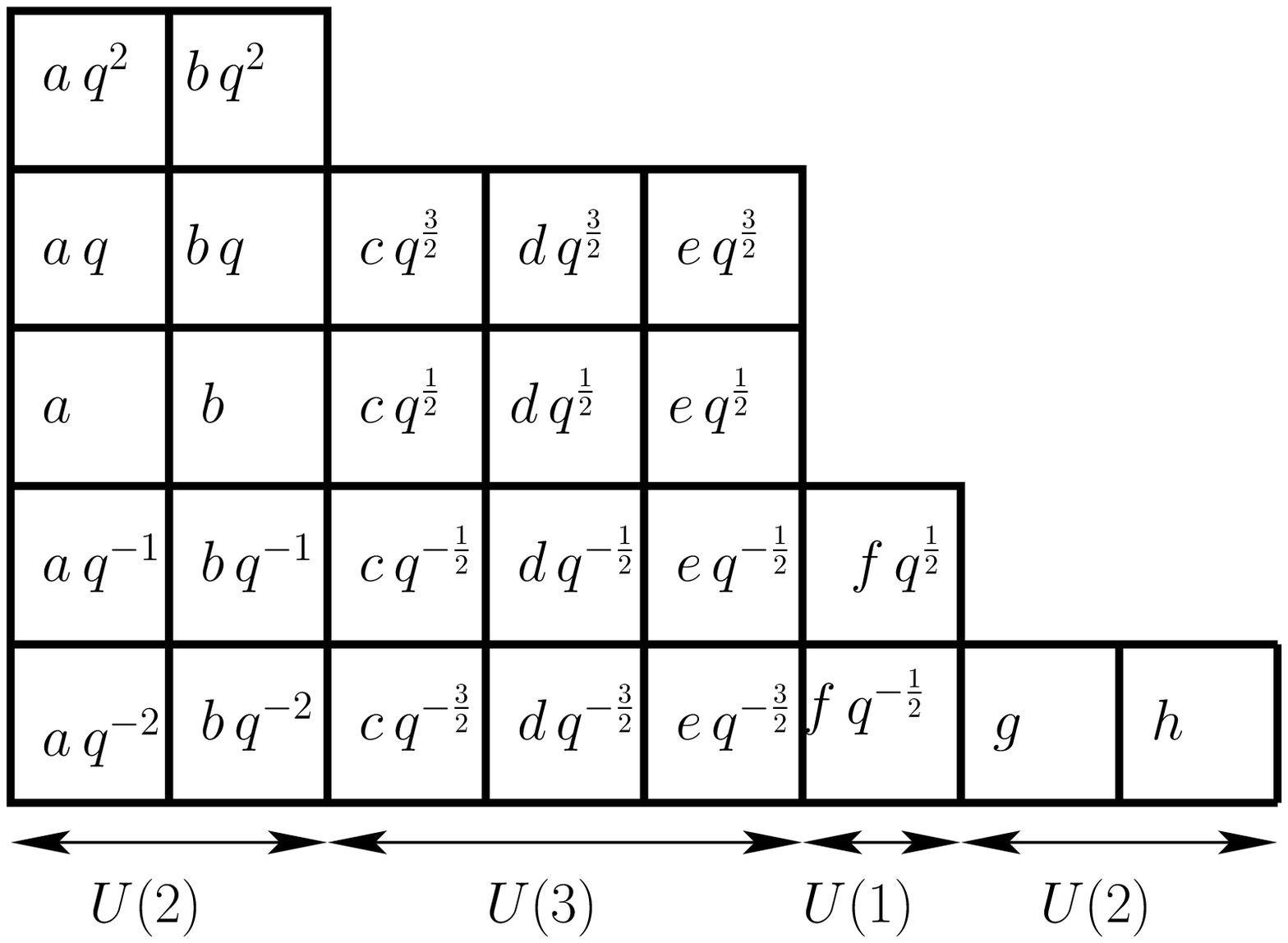}}
\end{center}
\begin{center}
\vspace{-.3in}
\caption{\label{YT} An example of the rule to associate flavor fugacities for  a non-maximal puncture. Illustrated here is a puncture for $N=26$
with
flavor symmetry $S(U(3)U(2)^{2}U(1))$. The $S(\dots)$ constraint imposes $(ab)^{5}(cde)^{4}f^{2}gh=1$.}
\end{center}
\end{figure}
\vspace{-.3in}
Generic punctures are classified~\cite{Gaiotto:2009we} by the embeddings   $SU(2)\subset SU(N)$, which are specified
 by the decomposition of the fundamental of  $SU(N)$ into $SU(2)$ representation.  (In the terminology of \cite{Chacaltana:2010ks},
 we focus on regular punctures).
 This information can be  encoded into a Young
 diagram with $N$ boxes, where the height of each column denotes the dimension of an $SU(2)$ representation. The commutant of this embedding
is the flavor symmetry associated to the puncture. The maximal puncture corresponds to a single-row diagram, the closed puncture ({\it i.e.} no puncture)
corresponds to a single-column diagram, and the $U(1)$ puncture to a two-column diagram with $N-1$ boxes in the first column and a single box in the second column.
The Young diagram in Fig.~\ref{YT} exemplifies a non-maximal
 puncture for $N=26$  with $S(U(3)U(2)^{2}U(1))$ flavor symmetry.
We are  lead  to the following conjecture for the index of a theory
with three generic punctures corresponding to Young diagarms
$\lambda_i$ \be &&{\mathcal I}(\Lambda_1,\Lambda_2,\Lambda_3)=
{\mathcal N}_{\lambda_1,\lambda_2,\lambda_3}(q)\,\prod_{i=1}^3 {\mathcal A}_{\lambda_i}(\Lambda_i)\times\nonumber\\
&&\qquad\sum_{{\mathcal R}}
  \,\frac{1}{{\rm dim}_q{\mathcal R}}\,
\chi_{{\mathcal R}}(\Lambda_1)\,\chi_{{\mathcal
R}}(\Lambda_2)\,\chi_{{\mathcal R}}(\Lambda_3)\,,\nonumber \ee with
$\Lambda_i$ labeling an association of  flavor fugacities according
to the Young diagram $\lambda_i$. The rule to associate the flavor
fugacities to the $SU(N)$ fugacities is illustrated in
Fig.~\ref{YT}. For all maximal punctures we have given the
normalization factors (${\mathcal N}$ and ${\mathcal A}$) above,
while for generic punctures these factors can be in principle
obtained by employing different S-dualities of the
quivers~\cite{Gaiotto:2009we}. As an example, consider the $E_7$
SCFT which is given by a sphere with two maximal punctures of
$SU(4)$ and one square Young diagram with four boxes. Following the
above procedure and fixing the normalization from the relevant
Argyres-Seiberg duality~\cite{Argyres:2007cn}, we are led to propose
\be &&{\mathcal
I}_{E_7}(\mathbf{x},\mathbf{y};a)=\frac{\exp\biggl[\sum_{n=1}^\infty\frac{q^{n}(1+q^n)}{1-q^{n}}\frac{a^{2n}+a^{-2n}}{n}\biggr]}{\eta^{\half}(\mathbf{x})\eta^{\half}(\mathbf{y})(1-q)(1-q^2)^2(1-q^3)
}\times\nonumber\\
&&\quad\sum_{{\mathcal R}}
  \,\frac{\chi_{{\mathcal R}}(\mathbf{x})\,\chi_{{\mathcal R}}(\mathbf{y})\,
\chi_{{\mathcal R}}(q^{\half}a,q^{-\half}a,q^{\half}/a,q^{-\half}/a)}{{\rm dim}_q{\mathcal R}}\,
\,,\nonumber
\ee Here $\mathbf{x},\,\mathbf{y}$ label the two sets of $SU(4)$ fugacities and $a$ the $SU(2)$ fugacity. The summation, as usual,
is over finite irreducible representations of $SU(4)$. We have verified perturbatively in $q$ that this expression is indeed $E_7$ covariant -- a tight check of our logic.

\section{Discussion}

We have given compelling evidence that the
reduced superconformal index of an ${\mathcal N}=2$ generalized $SU(N)$ quiver theory
is exactly computed by a correlator in $2d$  $SU(N)_q$ Yang-Mills.
This duality is new tool to investigate interacting field theories without a Lagrangian description.
For example, it should be useful to study the constraints obeyed by the Higgs branch operators,
generalizing to $N>3$ the analysis of \cite{Gaiotto:2008nz}.
Two-dimensional $q$YM  first appeared
 in a physical setting  in the context of counting BPS states~\cite{Aganagic:2004js},
and it would be interesting to find a relation with our work.
An obvious question is whether
our results  can be generalized to the full index, with all  fugacities turned on. It is already
remarkable that the {\it known} structure constants of the $SU(2)$ quivers implicitly define a $(q,p,u)$ deformation
of $SU(2)$ Yang-Mills. Work is in progress in investigating the  nature of this deformation,
 in order to  extrapolate it to $N >2$. The $q$ and $p$ fugacities appear on a symmetric footing, in a way
 which is strongly suggestive of an elliptic, or ``dynamical'',
 deformation of the quantum group structure $SU(N)_q$ that we have uncovered for $p=q$, $u=1$.
Indeed the full index is most elegantly expressed  \cite{Dolan:2008qi}  in terms of elliptic Gamma functions~\cite{Spiridonov}.
Finally, a more conceptual understanding of the duality would be very desirable. As for the AGT correspondence \cite{Alday:2009aq},
the existence, but not the details, of a $4d/2d$ relation can be traced to the definition
of the $4d$ SCFT as the infrared limit of the  $6d$ (2,0) theory on a Riemann surface.
Whether  this intuition can be turned into  a microscopic derivation remains to be seen.

\smallskip

\noindent{\bf Acknowledgments}:~
We would like to thank C.~Beem, D.~Gaiotto, S.~Gukov,  N.~Nekrasov and especially M. Aganagic and G. Moore for very useful discussions and suggestions.
This work was supported in part by DOE grant DEFG-0292-ER40697 and by NSF grant PHY-0969739.

\end{document}